\documentclass[12pt]{article}
\usepackage{amssymb,latexsym,epsfig,cite}

\newcommand{\be}{\begin{equation}}
\newcommand{\en}{\end{equation}}

\newcommand{\Sc}{Schr\"odinger}
\newcommand{\E}{equation}
\newcommand{\al}{\alpha }
\newcommand{\vfi}{\varphi }
\newcommand{\gm}{\gamma }

\def\bull{\vrule height 2ex width 1.5ex depth -.1ex }

\begin{document}

\thispagestyle{empty}

\baselineskip=15pt

\begin{center}
{\Large \bf Nonlocal SUSY deformations of\\ [1ex] 
periodic potentials}
\end{center}

\vskip1cm

\begin{center}
{David J. Fern\'andez C.$^*$, Bogdan Mielnik$^{*\dagger}$,\\[1ex]
Oscar Rosas-Ortiz$^*$, and Boris F. Samsonov$^\ddagger $}
\end{center}
\noindent
{\footnotesize
$^*$ Departamento de F\'{\i}sica, CINVESTAV-IPN, A.P. 14-740,
07000 M\'exico D.F., Mexico \\ [1ex]
$^\dagger$ Institute of Theoretical Physics, UW, Ho\.{z}a 69, Warsaw,
Poland\\ [1ex]
$^\ddagger $ Department of Quantum Field Theory, Tomsk State
University 36 Lenin Ave., 634050 Tomsk, Russia }

\vskip1cm
\begin{minipage}{13cm}

{\footnotesize
{\bf Abstract.} Irreducible $2$-order Darboux transformations are applied
to the periodic Schr\"odinger's operators. It is shown that for the pairs
of factorization energies inside of the same forbidden band they can
create new non-singular potentials with periodicity defects and bound
states embedded into the spectral gaps. The method is applied to the
Lam\'e and periodic piece-wise transparent potentials.  An interesting
phenomenon of {\it translational Darboux invariance} reveals nonlocal
aspects of the supersymmetric deformations.}
\end{minipage}

\vskip3cm
\hfill

\noindent 
{\small 
[1ex]

\newpage

\section{Introduction}

The one dimensional periodic potentials play an essential role in physics. 
The simple model of Kronig-Penney has been recently used for describing
superlattice structure of ultrathin epitaxial layers with diverse
constituent-semiconductor compositions such as ${\rm GaAs}$ and ${\rm
Al}_x{\rm Ga}_{1-x}{\rm As}$ \cite{PF}. The Kronig-Penney lattice with $R$
matrix interaction at the end of each block has been proposed for
analyzing the behavior of neutrons in crystals \cite{MM}. The Mathieu
equation has successfully described the diffraction of intense standing
light by strong sinusoidal media \cite{HJZ}.

The progress of these models is delayed by the fact that few analytically
solvable periodic potentials are known, including the text-book examples
of Kronig-Penney \cite{KrPe}, Lam\'e, and Mathieu \cite{BeEr}. The inverse
Sturm-Liouville methods could be applied \cite{L} but up to the authors
knowledge no simple periodic potentials have been found in this way. In
order to enlarge the class, the Darboux transformations of various orders
can be used \cite{DF,KS,crum,sukumar,AICD}, the subject commonly known as
the {\it supersymmetric quantum mechanics} (SUSY QM) 
\cite{ju96,ba01,cooper}. By employing 2-nd order irreducible techniques,
proposed initially for non-periodic potentials \cite{SMPLA} (see also a
hint by Krein \cite{krein}), the factorization energies above the bound
states can give new non-singular potentials \cite{FNN,aoyama,Trl}. We
shall now extend the method by showing that it can be applied in a
non-singular way to the periodic potentials if the $E$-values are in the
same spectral gap. In the most interesting cases, the operation can
generate the defects of the periodic structure, creating the bound states
inserted into spectral gaps (a typical phenomenon in solid state physics
\cite{blinowski}). 

Below, the technique will be applied to Lam\'e and periodically continued
soliton potentials. In its most inconspicuous form, it produces an
interesting non-local effect: the transformed potential becomes an exact
or approximate displaced copy of the initial one, a phenomenon which we
call the {\it translational invariance with respect to Darboux
transformations} or shortly {\it Darboux invariance} \cite{FMRS}. Quite
significantly, the effect shows itself asymptotically even if the
periodicity of the initial potential is not preserved. It permits to
understand the structure of Darboux generated lattice impurities as the
{\it contact effects} caused by a conflict between two non-local {\it
SUSY} transformations. 

\section{First and second order Darboux \\ transformations}

Let us outline briefly the main points of the Darboux method.  Consider
the \Sc \ \E \ with an arbitrary potential $V_0(x)$: 
\be
\begin{array}{ll} h_0\psi_E(x)=E\psi_E (x), & h_0=-\partial _x^2+V_0(x).
\end{array} \label{E0}
\en
The method permits to use the solutions $\psi_E(x)$ of the initial \Sc \
\E \ (\ref{E0}) to obtain the solutions $\vfi_E(x)$ of the transformed \Sc
\ \E \
\be \label{h1}
h_1\vfi_E=E\vfi_E, \quad h_1=-\partial _x^2+V_1(x)
\en
by applying a certain differential operator $L$, $\vfi_E(x)=L\psi_E(x)$. 
In the first order case $L=-\partial_x+w(x)$, the transformation yields: 
\be\label{DV}
\Delta V(x)=V_1(x)-V_0(x)=-2\left[\ln u(x)\right]''
\en
where $u(x)$ is a solution of (\ref{E0}) called the {\it transformation
function} while $w(x)=[\ln u(x)]'$ is the {\it superpotential}. Note that
the method applies for any $E \in {\mathbb R}$ in the spectrum or in the
resolvent set of $h_0$; it links the solutions of (\ref{E0}) and
(\ref{h1}), without requiring that $\psi_E$ should be bounded or square
integrable (though in some problems, such assumptions can be pertinent).
Quite obviously, in order to avoid the creation of singularities in $w(x)$
and in $V_1(x)$ one must look for nodeless $u(x)$. 

In the second order case, the new potentials and eigenfunctions are
defined by a pair of transformation functions $u_1(x)$ and $u_2(x)$,
$h_0u_{1,2}=\alpha_{1,2}u_{1,2}$ (we use deliberately the symbols
$\alpha_1, \ \alpha_2$ instead of $E_1, \ E_2$ to stress that both
parameters do not need to belong to the spectrum of $h_0$). The
transformation reads: 
\begin{equation}
\begin{array}{cl}
\vfi_E= & W^{-1}(u_1,u_2)W(u_1,u_2,\psi_E), \nonumber \\
\Delta V=& -2[\ln W(u_1,u_2)]'' \label{Wr}
\end{array}
\end{equation}
where $W$ denotes the Wronskian of the corresponding functions. 

For the periodic potentials $V_0(x) \equiv V_0(x + T)$ the method is often
affected by singularities. Some authors opt to use the band edge solutions
\cite{DF,KS,FNN} (which are periodic or antiperiodic \cite{RS}).  We are
going to show that the use of the generalized {\it Bloch functions}
belonging to the same spectral gap brings even more interesting results.

\section{Bloch eigenfunctions and Darboux \\ transformations}

The Bloch functions are usually defined as physically interpretable
eigenfunctions of (\ref{E0}) for $E$ belonging to the spectrum of $h_0$.
Yet, they exist also out of the spectral area. By writing (\ref{E0}) for
any $E\in{\mathbb R}$ as the first order differential equation for a
vector formed by $\psi$ and its derivative $\psi'$
\begin{equation}\label{5}
\frac{d}{dx}
\left[
\begin{array}{c}
\psi \\ \psi'
\end{array}
\right]
= \left[
\begin{array}{cc}
0 & 1 \\ V_0-E & 0
\end{array}
\right]
\left[ \begin{array}{c}
\psi \\ \psi'
\end{array}
\right]
\end{equation}
one sees that $\psi(x) , \ \psi'(x)$ are given by a certain linear
transformation applied to $\psi(0), \ \psi'(0)$:
\begin{equation}
\left[ \begin{array}{c}
\psi(x) \\ \psi'(x)
\end{array} \right] =
b(x)\left[ \begin{array}{c}
\psi(0) \\ \psi'(0)
\end{array}
\right] \label{trans}
\end{equation}
where $b(x)$ is a $2\times 2$ simplectic {\it transfer matrix}
\cite{RS,bmma}. Note that $b(x)$ can be expressed in terms of special real
solutions ${\rm v}_1(x)$, ${\rm v}_2(x)$ of (\ref{5})  defined by the
initial data ${\rm v}_1(0)=1$, ${\rm v}_1'(0)=0$, ${\rm v}_2(0)=0$, ${\rm
v}_2'(0)=1$: 
\begin{equation}
b(x) = \left( {\begin{array}{*{20}c}
  {{\rm v}_1 (x)} & {{\rm v}_2 (x)}  \\
   {{\rm v}_1 ^\prime  (x)} & {{\rm v}_2 ^\prime (x)}  \\
\end{array}} \right),
\end{equation}
implying ${\rm Det}[b(x)] \equiv 1$. If $V_0(x)$ is periodic, an essential
role belongs to the {\it Floquet} (or {\it monodromy}) {\it matrix} \
$b(T)$ (see e.g.  \cite{RS,fraboma}). Since ${\rm Det}[b(T)]=1$, its
eigenvalues $\beta$ are given by: 
\be\label{CC}
\beta ^2 - D\beta + 1 = 0
\en
where
\begin{equation}
D = D(E) = {\rm Tr} [b(T)] \label{discriminant}
\end{equation}
is called the {\it Lyapunov function, Hill determinant} or {\it
discriminant} of (\ref{E0})  \cite{L,RS,MW}. As follows from (\ref{CC}),
$\beta $ takes two values $\beta_{\pm}$ such that $\beta_+ \beta_- =1$: 
\be\label{be} 
\beta_{\pm}=D/2\pm \sqrt {D^2/4-1}\ .  
\en 
Whenever for any $E\in {\mathbb R}$, one of the eigenvectors of $b(T)$
(for either $\beta = \beta_+$ or $\beta = \beta_-$) is used as an initial
condition for $\psi$ and $\psi'$ at $x=0$, it originates a special
solution of (\ref{E0}) for which the transfer law (\ref{trans}) at $x=T$
reduces to: 
\begin{equation} \psi(T) = \beta \psi(0),
\quad \psi'(T)  = \beta \psi'(0)  \label{bloch} \end{equation} 
and more generally
\begin{equation} 
\hskip-0.5cm \psi(x+nT) = \beta^n \psi(x), \ \psi'(x+nT)
= \beta^n \psi'(x), \label{blochit} 
\end{equation} $n=0, \pm 1, \dots$ 
The eigenfunctions $\psi(x)$ of (\ref{E0}) which fulfill
(\ref{bloch},\ref{blochit}) exist for any parameter $E=\alpha \in {\mathbb
R}$ (not necessarily belonging to the energy spectrum of $h_0$)  and are
called the {\it Bloch functions}. Their structure depends essentially on
the values of $E$. 

If $|D(E)| < 2$, $E$ is {\it inside} of a spectral band and the
eigenvalues $\beta_{\pm}$ in (\ref{be}) are complex numbers of modulus
$1$. We can put then
\be\label{DE}
\beta_+=\exp(ikT)  \quad   \beta_-=\exp(-ikT)
\en
where $k$ is a real parameter called the {\it crystal quasimomentum}.  The
equations (\ref{be})  and (\ref{DE}) define an implicit function $k=k(E)$
called the {\it dispersion law}. The corresponding bounded and essentially
complex Bloch functions are the ones traditionally considered in the solid
state physics.

The equation $|D(E)|=2$, in turn, defines the band edges \cite{RS}. 
Following \cite{FNN} let us denote them by
$$
E_0<E_1\le E_{1'}<E_2\le E_{2'}<\ldots <E_j\le E_{j'}<\ldots
$$
At each band edge the Bloch eigenvalues (\ref{be}) are $\beta_+ = \beta_-
= \pm 1$, and the degenerate $b(T)$ defines one Bloch function, periodic
or antiperiodic. Both edge eigenfunctions $\psi_{j}$ and $\psi_{{j'}}$ are
real and have the same number $j$ of nodes. 

If $|D(E)|> 2$, the eigenvalues (\ref{be}) are real $\beta_+ = \beta, \
\beta_- = 1/\beta$ ($0\neq \beta \in {\mathbb R}$);  $b(T)$ has the real
eigenvectors for both $\beta=\beta_\pm$, originating the real Bloch
functions, which form a natural basis for the general solutions of
(\ref{E0}). (In what follows, when speaking about the Bloch functions in
this regime, we shall always have in mind the {\it real Bloch functions},
without specially mentioning it. The complex Bloch functions \cite{WV}
have a constant phase and are easily reduced to the real ones.) Note that
all solutions of (\ref{E0}) diverge in either $+\infty$ or $-\infty$,
hence $E$ is in a forbidden band (resolvent set of $h_0$). The Bloch
functions here are deprived of a physical meaning but happen to be crucial
as transformation functions in the Darboux algorithms. Their properties
still depend on the specific localization of $E$ in the resolvent set.

The interval $(-\infty,E_0)$ constitutes the lowest forbidden band where
$D>2$ and $\beta_{\pm}>0$. In the following spectral gaps $(E_j,E_{j'})$
one has either $D>2$ and $\beta_{\pm}>0$ for $j$ even or $D<-2$ and
$\beta_{\pm}<0$ for $j$ odd.

Let $u(x)$ be one of the Bloch eigenfunctions for $E=\alpha<E_0$. Due to
(\ref{blochit}), $u(x)$ has either no zeroes at all or a periodically
distributed infinite set. The last possibility cannot occur since then,
according to the Sturm oscillator theorem (see e.g.  \cite{Ber}), the
ground state eigenfunction should have an infinity of zeroes as well,
which contradicts the fact that $\psi_{0}(x)$ is nodeless. Thus, for any
$\alpha<E_0$ there exist two linearly independent {\it nodeless} Bloch
eigenfunctions, both suitable to generate non-singular Darboux
transformations (\ref{DV}). As nodeless must be the Wronskian $W(u_1,u_2)$
for any two Bloch functions $u_1, \ u_2$ with $\alpha_1, \alpha_2 < E_0$,
thus generating a non-singular reducible transformation (\ref{Wr}) 
\cite{BS}.

Let us now examine the situation in the higher spectral gaps. We shall
show that while the first order Darboux transformations here are singular,
the second order ones can be regular, in analogy with \cite{S}. Indeed one
has: 

{\bf Proposition 1.} Let $u_1, \ u_2$ be two linearly independent Bloch
eigenfunctions of (\ref{E0}), $u_i(x+T) = \beta_i u_i(x), \ h_0 u_i =
\alpha_i u_i$, with $\alpha_1, \ \alpha_2$ in the spectral gap $(E_j,
E_{j'})$ of a continuous, periodic $V_0(x)$. Then: (i) In each periodicity
interval $[x_0,x_0+T)$ both $u_1(x)$ and $u_2(x)$ have $j$ roots,
depending continuously on $\alpha_1$ and $\alpha_2$;  (ii) if $\alpha_1
\neq \alpha_2$, or $\alpha_1 = \alpha_2 = \alpha$ but $\beta_1 \neq
\beta_2$, then the roots of $u_1$ and $u_2$ cannot coincide; they form two
infinite alternating sequences extending from $-\infty$ to $+\infty$; 
(iii) the Wronskian $W(x) \equiv W(u_1,u_2)$ has no nodal points on
${\mathbb R}$ (and so, defines a non-singular transformation (\ref{Wr})). 

{\bf Proof.} (i) is well known in the theory of Hill's equations
\cite{MW}. (ii) If $\alpha_1 < \alpha_2$ then due to the oscillatory
theorem \cite{Ber}, between each two neighbouring nodes of $u_1$ there is
at least one nodal point of $u_2$ (but no more than one because of (i)).
Moreover, the nodal point of $u_1$ cannot be nodal for $u_2$, since then
the oscillatory theorem would require still one more node of $u_2$
in between the vicinal nodes of $u_1$; so $u_2$ would have more nodes than
$u_1$ in $[x_0,x_0 +T)$. If $\alpha_1 = \alpha_2$ but $\beta_1 \neq
\beta_2$, $u_1, \ u_2$ cannot have a common nodal point. If they did, they
would satisfy proportional initial conditions at the common node and so,
they would be proportional all over ${\mathbb R}$, which is impossible if
$\beta_1 \neq \beta_2$. Finally, due to (\ref{blochit}) the zeroes of
$u_1$ and $u_2$ are periodically distributed on ${\mathbb R}$, so they
must form two alternating sequences extending from $-\infty$ to $+\infty$. 
(iii) Let now $\nu$ and $\tilde\nu$ be two neighbouring nodes of $u_1(x),
\ u_2(x)$ respectively. Observe, that at the extremes of the interval
$(\nu,\tilde\nu)$ the values $W(\nu) = - u_2(\nu)u_1'(\nu)$ and
$W(\tilde\nu)  = u_1(\tilde\nu)  u_2'(\tilde\nu)$ cannot vanish and must
have the same sign. Indeed, since we assume that neither $u_1$ nor $u_2$
have zeros inside of $(\nu,\tilde\nu)$ the sign of $u_1(x)$ coincides with
the sign of $u_1'(\nu)$ while the sign of $u_2(x)$ is opposite to that of
$u_2'(\tilde\nu)$ in all $[\nu,\tilde\nu]$. Moreover, the first derivative
$W'(u_1,u_2) = (\alpha_1 - \alpha_2) u_1 u_2$ either vanishes everywhere
(if $\alpha_1 = \alpha_2$), or nowhere in $(\nu,\tilde\nu)$ (if $\alpha_1
\neq \alpha_2$). In both cases $W(u_1,u_2)$ is monotonic, extends between
two non-zero values of the same sign and so, it cannot vanish in
$[\nu,\tilde\nu]$. Since the same holds for any other neighbouring nodes
of $u_1$ and $u_2$ then $W(u_1,u_2)$ has no nodal points on the entire
${\mathbb R}$.  \bull

\medskip

\begin{figure}[ht]
\centering \epsfig{file=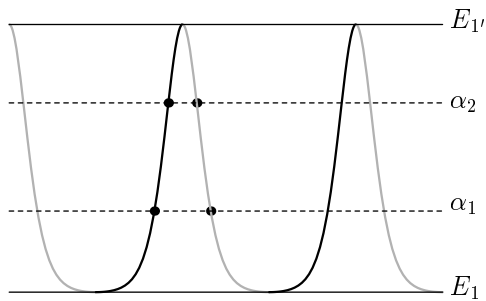, width=7cm}
\caption{\footnotesize The ``nodal curves'' illustrating the
$\alpha$-dependence of the nodal points of two Bloch functions $u^\beta$
(black line) and $u^{1/\beta}$ (shadow line). Notice that any two vicinal
nodes situated at $\alpha = \alpha_1$ are never separated just by a single
node at $\alpha = \alpha_2$, thought they can be separated by a pair. The
graphic was obtained for the Lam\'e potential (\ref{lame}) with $n=1, \
m=0.99$.} 
\end{figure} 

While for a fixed $\alpha \in (E_j, E_{j'})$ the zeroes of the Bloch pair
are isolated points, when $\alpha$ varies, they draw a sequence of
continuous {\it nodal curves}. These curves cannot intersect inside of
$(E_j,E_{j'})$ and, moreover, each one can intersect any vertical line
$x={\rm c}$ only once (compare Proposition~1ii), so they form a sequence
of strictly monotonic branches which meet at the band edges, where two
Bloch solutions degenerate to one (see Fig.~1); a pattern which permits to
extend the Proposition~1 to non-trivial linear combinations of the Bloch
functions. In what follows we shall use the simplified symbol $u^\beta$ to
denote the Bloch eigenfunction for any $\alpha$, corresponding to the
Bloch eigenvalue $\beta$ (so to be exact $u^\beta \equiv
u(x,\alpha,\beta)$).

\medskip

{\bf Theorem 1.} Let $u_1(x) = u^{\beta_1}(x), \ u_2(x) = u^{\beta_2}(x)$
be two nontrivial Bloch functions of $h_0$ for two different eigenvalues
$\alpha_1, \alpha_2 \in (E_j, E_{j'})$ and let $v_1(x)  = u^{\beta_1} +
\kappa_1 u^{1/\beta_1}$, $v_2(x) = u^{\beta_2} + \kappa_2 u^{1/\beta_2}$,
where $(\kappa_1,\kappa_2)\in {\mathbb R}^2$. Then there exists an
infinite closed sector in ${\mathbb R}^2$, bordered by $\kappa_1$- and
$\kappa_2$-axes, where the Wronskians $W(v_1,v_2)$ are nodeless. 

{\bf Proof.} Note that the Bloch functions $u^{\beta_1},u^{1/\beta_1} \
(\alpha = \alpha_1)$ and $u^{\beta_2},u^{1/\beta_2} \ (\alpha = \alpha_2)$
determine four sequences of nodal points on ${\mathbb R}$. In view of the
nodal configuration of Fig.~1, each two neighbouring nodes
$\nu_1,\tilde\nu_1$ of $u^{\beta_1}$ and $u^{1/\beta_1}$ can be separated
by a pair of nodes $\nu_2,\tilde\nu_2$ of $u^{\beta_2}$, $u^{1/\beta_2}$
but not by just one of them.  Without loosing generality one can choose
the point $x=0$ so that $u^{\beta_i}(0) \neq 0$, $u^{1/\beta_i}(0) \neq 0$
and that the four subsequent nodal points are $\nu_1, \tilde \nu_1, \nu_2,
\tilde \nu_2$ (belonging to $u^{\beta_1}, \ u^{1/\beta_1}, \ u^{\beta_2},
\ u^{1/\beta_2}$ respectively). By multiplying each Bloch functions by
$\pm 1$ one can also achieve $u^{\beta_i}(x_0) >0$ and $u^{1/\beta_i}(x_0) 
>0$. According to Proposition~1, all Wronskians
$W(u^{\beta_1},u^{\beta_2}), \ W(u^{\beta_1},u^{1/\beta_2}), \
W(u^{1/\beta_1},u^{\beta_2}), \ W(u^{1/\beta_1},u^{1/\beta_2})$ have
constant signs; moreover, it is straightforward to check that they are now
strictly positive. Henceforth, by choosing $\kappa_1, \kappa_2 \geq 0$ one
obtains
$$
W(v_1,v_2) = W(u^{\beta_1},u^{\beta_2}) + \kappa_2
W(u^{\beta_1},u^{1/\beta_2}) + \kappa_1 W(u^{1/\beta_1},u^{\beta_2}) +
\kappa_1 \kappa_2 W(u^{1/\beta_1},u^{1/\beta_2}) 
$$ 
strictly positive, producing a non-singular Darboux transformation
(\ref{Wr}). \bull


\section{The translational effects of the Darboux \\ operations} 

We shall show now that for certain classes of potentials the Darboux
transformations might yield an interesting non-local effect. Consider
first of all the 1-soliton well \cite{matveev}
\be
\label{solit}
V_0(x)=-2\gm_0 ^2{\rm sech} ^2\gm_0 x\ ,\quad  \gm_0>0,
\en
with one eigenvalue $E_0=-\gm_0 ^2$.  We shall check that the wells
(\ref{solit}) admit special Darboux transformations leading to the exact
coordinate displacements. Note that the Schr\"odinger equation (\ref{E0})
with the potential (\ref{solit}) can be exactly solved for any
$E\in{\mathbb R}$ by applying the Darboux transformation (\ref{DV}) to the
free Hamiltonian. One of the solutions is: 
\begin{equation}
u(x) = \frac{\cosh\gamma_0(x+\delta_1)}{\cosh\gamma_0 x} e^{-\gamma_1 x}
\label{eigen}
\end{equation}
If now $\alpha_1 = - \gamma_1^2<E_0$, then using (\ref{eigen}) as a 1-SUSY
transformation function for our 1-soliton well $V_0(x)$, one obtains a
displaced version of (\ref{solit}), $V_1(x) = V_0(x + \delta_1)$, with
\begin{equation} 
\delta_1 = \frac{1}{\gamma_0}{\rm artanh} 
\frac{\gamma_0}{\gamma_1} 
\end{equation} 

If $\alpha_1 >E_0$, the first order Darboux transformation can neither
produce the displaced, nor even nonsingular potential, but the second
order SUSY opens new possibilities. Taking $\delta_1 = \delta + i
\frac{\pi}{2\gamma_0} = \delta + i \tau'$, where $\tau'$ is half the
imaginary period of the transparent well (\ref{solit}), we easily induce a
{\it complex} displacement generated by $u_1 = e^{-\gamma_1 x}
\sinh\gamma_0(x+\delta)/\cosh\gamma_0 x$ and leading to the {\it real but
singular} $V_1(x) = 2 \gamma_0^2 {\rm csch}^2\gamma_0 x$. By repeating now
the operation with a new complex $\delta_2 = \delta' - i
\frac{\pi}{2\gamma_0}$ one returns to the original transparent well
displaced additionally by $\delta'' = \delta_1 + \delta_2 = \delta +
\delta'$. The corresponding nodeless Wronskian in (\ref{Wr}) is: 
\begin{equation} 
W(u_1,u_2) =
(\gamma_1 - \gamma_2)\frac{\cosh\gamma_0 (x + \delta'')}{ \cosh\gamma_0
x}e^{-(\gamma_1 + \gamma_2)x}. \label{w1sol}
\end{equation} 

Until now our second order displacements (\ref{Wr}) were backed by pairs
of first order transformations (\ref{DV}) of the displacement type, with
real or complex $\delta$'s. It would be interesting to achieve the same
effect without this kind of first order scenario. This indeed happens for
the symmetric 2-soliton well
\begin{equation}
V_0(x)=\frac{2(\gm_1^2-\gm_2^2)(\gm_1^2 {\rm sech}^2\gm_1 x + \gm_2^2 {\rm
csch}^2\gm_2 x)}{(\gm_1\tanh\gm_1 x - \gm_2\coth\gm_2 x)^2} \label{2sol}
\end{equation}
obtained from the null potential using (\ref{Wr}) with $u_1(x) =
\cosh\gm_1 x, \ u_2(x)  = \sinh\gm_2 x,$ where $\gm_2>\gm_1$. The
potential (\ref{2sol}) has two discrete energy levels at $E_0 = -\gm_2^2,
\ E_1 = - \gm_1^2$. Denote now $u_3(x) = e^{-\gm_3 x}, u_4(x) = e^{-\gm_4
x}$ and $W_{ij}(x) \equiv W(u_i(x), u_j(x))$, $W_{ijl}(x) \equiv W(u_i(x),
u_j(x),u_l(x))$; then apply an isospectral second order Darboux
transformation to the potential (\ref{2sol}) using two new eigenfunctions
$\tilde u_3(x)  = W_{123}(x)/W_{12}(x), \ \tilde u_4(x) =
W_{124}(x)/W_{12}(x)$. A straightforward calculation shows:
\begin{equation}
\tilde W_{34}(x) = e^{-(\gm_3+\gm_4)x}(\gm_3 - \gm_4)\Gamma \,
W_{12}(x+\delta)/W_{12}(x) \label{2displ}
\end{equation}
where
\begin{eqnarray}
& \Gamma^2 =  (\gm_1^2 - \gm_3^2)(\gm_2^2 - \gm_3^2)(\gm_1^2 -
\gm_4^2)(\gm_2^2 - \gm_4^2) \label{ddc}
\end{eqnarray}
As one can easily see, $\tilde W_{34}(x)$ induces a second order Darboux
displacement, $V_1(x) = V_0(x+\delta)$, if and only if the following two
numbers coincide:
\begin{eqnarray} && \delta =
\frac{1}{\gm_1} {\rm artanh}\left[ \frac{\gm_1(\gm_3 + \gm_4)}{\gm_1^2 +
\gm_3 \gm_4}\right] \label{dda} \\ && \delta = \frac{1}{\gm_2} {\rm
artanh}\left[\frac{\gm_2(\gm_3 + \gm_4)}{\gm_2^2 + \gm_3\gm_4}\right]
\label{ddb} 
\end{eqnarray} 
Notice that there are three regions in which (\ref{2displ}) yields a
non-singular Darboux transformation (\ref{Wr}), $\Omega_1=\{
\gm_3,\gm_4<\gm_1\}, \ \Omega_2=\{ \gm_1<\gm_3,\gm_4<\gm_2\}, \ \Omega_3
=\{ \gm_3,\gm_4>\gm_2\}$, but only in $\Omega_2$ can one achieve the
consistency between (\ref{dda}) and (\ref{ddb}), visualized by the
intersection of two surfaces represented on Fig.~2. The points
$(\gm_3,\gm_4)$ on the intersection curve provide the precise data for the
displacement, with $\delta$ defined as the common value of (\ref{dda}) and
(\ref{ddb}).

\begin{figure}[ht]
\centering \epsfig{file=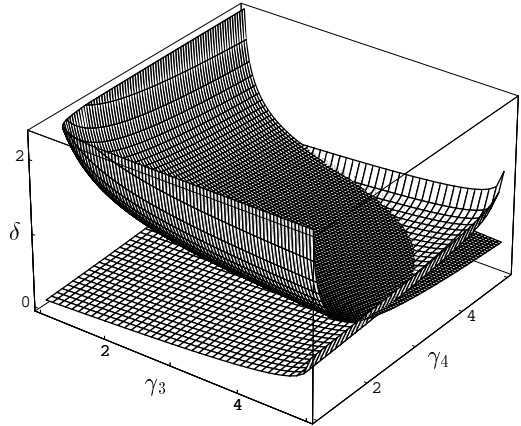, width=7cm}
\caption{\footnotesize The intersection of two surfaces (\ref{dda}) and
(\ref{ddb}) provides the consistent data for the second order Darboux
displacements of the 2-solitonic potential (\ref{2sol}).}
\end{figure}

The approximate forms of all these phenomena can be observed for the
``collage potentials'' composed of fragments of the transparent wells. The
simplest case is the periodic potential obtained by truncating
(\ref{solit}) to a finite interval $[-a,a]$ and then periodically
repeating all over ${\mathbb R}$. One arrives at a periodic $V(x)$ with
$T=2a$, for which the Schr\"odinger equation (\ref{E0}) can be explicitly
solved for any $E=k^2$, permitting to calculate the Lyapunov function
\cite{EJP}
\begin{eqnarray}
\hskip-0.8cm & \frac 12D=
\frac {w_0}k \left(\frac {w_0^2 - 2k^2-\gm_0 ^2}{k^2+\gm_0 ^2} \right) 
\, \sin 2ka\
\vphantom{\int _{\frac 22}^\frac 22}
+\left( 1-\frac{2w_0^2}{k^2+\gm_0 ^2} \right) \cos 2ka 
\vphantom{\int _{\frac 22}^\frac 22} \nonumber
\end{eqnarray}
(with $w_0=\gamma_0\tanh\gamma_0 a$). Notice that the equation defining
the band edges $\vert D\vert = 2$ is no more complicated than for the
Kronig-Penney potential. The discrete energy level of the original
potential (\ref{solit}) at $E_0= -\gm_0^2$ belongs now to the lowest
allowed band \cite{EJP}. The former ground level expands into the first
spectral band and the interval $(E_0, \infty)$ splits into the subsequent
gaps and bands.

\medskip
\begin{figure}[ht]
\centering \epsfig{file=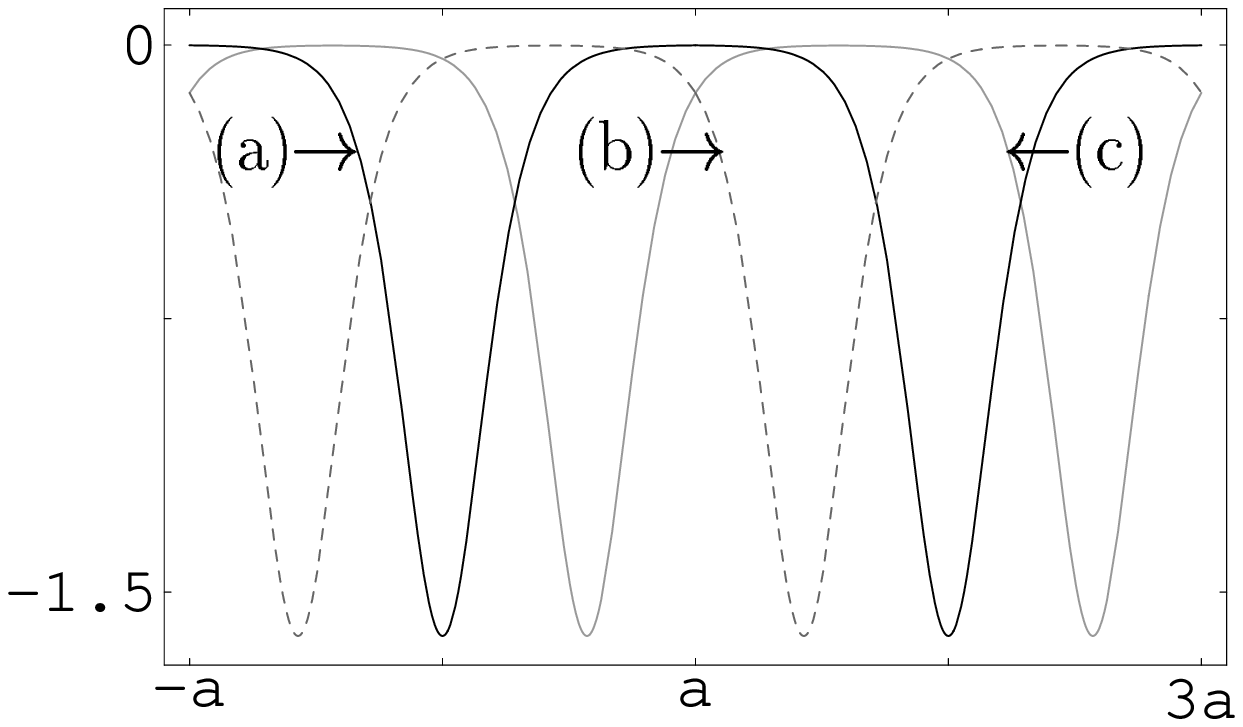, width=7cm}
\caption{\footnotesize Second order Darboux transformations of the
periodically continued one soliton potential with $a=5$ and
$\gamma_0=0.9$.  (a) The initial potential; (b-c) The modified forms after
using (\ref{Wr})  with $u_1, \ u_2$ chosen to be the pairs of Bloch
functions for $\al_1=-2$ and $\al_2=-0.9$.}
\end{figure}
\medskip

To illustrate the result of the Darboux transformation, we choose $a=5$
and $\gm_0 =0.9$, the numerical values of the lowest band edges becoming
$E_0=-0.8107$, $E_1=-0.8090$, $E_{1'}=0.0001$, $E_2=0.1578$,
$E_{2'}=0.1580$;  $E_3=0.5926$, $E_{3'}=0.5929$. If now the transformation
(\ref{Wr}) is applied, the transformed potential $V_1(x)$ approximates
very well a displaced copy of $V(x)$: the effect looks as if the main body
of $V_0(x)$ was displaced, leaving only some tiny remnants in vicinities
of the former peaks (Fig.3). 

\begin{figure}[htbp]
\centering \epsfig{file=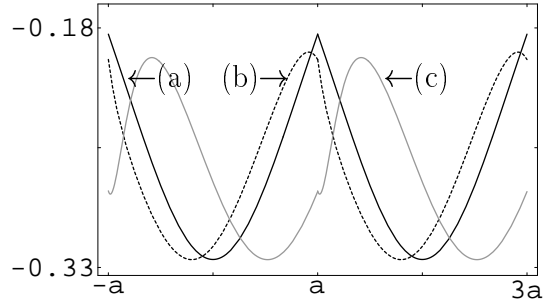, width=7.5cm}
\caption{\footnotesize Second order Darboux transformations of the
periodically continued one soliton potential with $a=2$ and
$\gamma_0=0.4$.  (a) The initial potential; (b-c) The modified forms after
using (\ref{Wr})  with $u_1, \ u_2$ chosen to be the pairs of Bloch
functions for $\al_1=-10$ and $\al_2=-2$.}
\end{figure}
\medskip

The phenomenon quite obviously imitates the behaviour of the original
1-soliton well, with an accuracy depending on the size of the repeated
1-soliton fragment (the bigger the fragment the better the effect). To
examine this accuracy, we have performed the next computer experiment for
the periodic potential composed of smaller pieces of (\ref{solit}), taking
$\gamma_0 =0.4$ and $a=2$. The lowest band edges are now $E_0 =-0.2664$,
$E_1= 0.3204$, $E_{1'}=0.3817$, $E_2=2.1967$, $E_{2'}=2.2073$,
$E_3=5.2838$, $E_{3'}=5.2885$, and the results of the 2-order Darboux
transformation are shown on Fig.4. One can again observe a displacement
affecting the main part of the original potential, though now the points
of non-differentiability visibly resist the operation. 

An even more interesting effect occurs for the fragmented 2-soliton wells. 
By choosing the truncation borders $\pm a$ exactly at its minima, then
repeating periodically and applying (\ref{Wr}) one gets a surprisingly
exact picture of the displacement (Fig.~5). 

\medskip

\begin{figure}[htbp]
\centering \epsfig{file=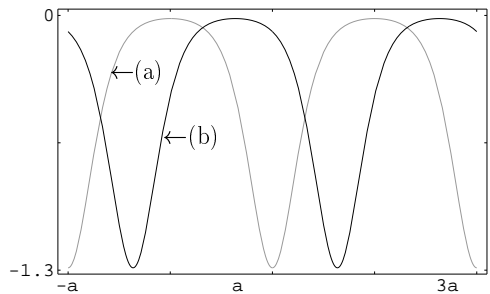, width=6.5cm}
\caption {\footnotesize Darboux operations on a periodically continued two
soliton potential with $\gm_1= 0.8$, $\gm_2= 0.805$ and $a=4.0279$. (a) 
The initial potential; (b)  the result of a 2-order Darboux transformation
for a pair of Bloch functions with $\al_1=-0.6$, $\al_2=-0.0007$ in the
first energy gap $(E_1=-0.6359,E_{1'}=0)$. Notice a very good
approximation to a finite displacement.}
\end{figure}
\medskip

The effect once again cannot be perfect (the points of discontinuity of
the third derivative are now fixed), but the difference between the
supersymmetrically transformed and displaced potential on our scale is
practically invisible. 

Significantly, the existence of this kind of non-local phenomena (exact or
approximate) tells a lot about the nature of the periodicity defects
which can be supersymmetrically generated. To see this, let us consider a
class of periodic potentials where the (non-local)  displacements appear
in their purest form. 

\section{Supersymmetric transformations of the \\ Lam\'e potentials} 

We refer to the Lam\'e potentials, frequently considered in
crystallography:
\be
V_0(x) = n(n+1) m \,\, {\rm sn}^2(x|m), \quad n\in{\mathbb N} \label{lame}
\en
where ${\rm sn}(x|m)$ is the standard Jacobi elliptic function. The
potentials (\ref{lame}) have exactly $2n + 1$ band edges, $n+1$ allowed
and $n+1$ forbidden bands, and there exist analytic formulae for the
eigenfunctions at the band edges (see e.g. \cite{FNN}).

We shall show that the global SUSY displacement is a typical phenomenon in
the subclass of Lam\'e functions with $n=1$, already for the first order
Darboux transformations. In fact, we have

{\bf Theorem 2.} The Bloch eigenfunctions $u^\beta(x), \ u^{1/\beta}(x)$
for a factorization energy $\alpha\leq E_0$ used in the first order
Darboux transformations generate a displacement $V_1(x) = V_0(x + \delta)$
of an arbitrary periodic $V_0(x)$ if and only if
\begin{equation}
u^\beta(x)u^{1/\beta}(x+\delta) = c \label{restriction}
\end{equation}
where $c$ is a constant.

{\bf Proof.} Suppose that $u^\beta(x)$ induces the Darboux displacement
$V_1(x) = V_0(x + \delta)$. The standard Darboux theory and $u^\beta(x +
T) = \beta u^\beta(x)$ imply that $\tilde u^{1/\beta}(x)  \propto
1/u^\beta(x)$ is the Bloch eigenfunction for $h_1$ with the same
factorization energy $\alpha$, so that $\tilde u^{1/\beta}(x+T)= (1/\beta)
\tilde u^{1/\beta}(x)$. As $V_1(x) = V_0(x+\delta)$, the coordinate change
$x \rightarrow x+\delta$ in the Schr\"odinger equation for $\tilde
u^{1/\beta}(x)$ and the fact that the Bloch eigenfunctions are unique up
to a multiplicative factor imply that $\tilde u^{1/\beta}(x)  \propto
u^{1/\beta}(x + \delta)$. Hence
\begin{equation}
u^{1/\beta}(x + \delta) = \frac{c}{u^\beta(x)} \ \Rightarrow \ u^\beta(x)
u^{1/\beta}(x + \delta) = c
\end{equation}
Conversely, suppose that the Bloch eigenfunctions of $h_0$ for the
eigenvalue $\alpha$ satisfy $u^\beta(x)  u^{1/\beta}(x + \delta) = c$. 
Then perform a first order Darboux transformation using $u^\beta(x)$. In
terms of $u^\beta(x)$ and $u^{1/\beta}(x)$ the initial potential is given
by: 
\begin{equation}
V_0(x) = \frac{[{u^\beta}(x)]''}{u^\beta(x)}+ \alpha =
\frac{[u^{1/\beta}(x)]''}{u^{1/\beta}(x)}+ \alpha .
\label{inpot}
\end{equation}
Quite similarly, the final potential can be expressed in terms of $\tilde
u^{1/\beta}(x) \propto 1/u^\beta(x) = u^{1/\beta}(x + \delta)/c$: 
\begin{equation}
V_1(x) = \frac{[\tilde u^{1/\beta}(x)]''}{\tilde u^{1/\beta}(x)}+
\alpha
= \frac{[u^{1/\beta}(x+\delta)]''}{u^{1/\beta}(x+\delta)}+
\alpha. \label{endpot}
\end{equation}
By comparing (\ref{inpot}) and (\ref{endpot}) one immediately sees: 
$V_1(x) = V_0(x+\delta)$. \bull

\medskip

Notice now that the criterion (\ref{restriction}) indeed holds for the
Lam\'e potentials with $n=1$. In order to prove that, consider the Bloch
functions associated to the corresponding Lam\'e equation (see \cite{WV},
Section 23.7): 
\begin{eqnarray}
&&\hskip-1cm u^\beta(x) =
\frac{\sigma(x_0+\omega')}{\sigma(x_0+a+\omega')}\frac{\sigma(x+a+\omega')}{ 
\sigma(x+\omega')}e^{-\zeta(a)(x-x_0)}\nonumber \\
&&\hskip-1cm u^{1/\beta}(x) =
\frac{\sigma(x_0+\omega')}{\sigma(x_0-a+\omega')} 
\frac{\sigma(x-a+\omega')}{
\sigma(x+\omega')}e^{\zeta(a)(x-x_0)} \label{blochlame}
\end{eqnarray}
where $x_0$ is a fixed point in $[0,T=2K)$ selected so that
$u^\beta(x_0)=u^{1/\beta}(x_0) =1$, $\beta = \exp[2a\zeta(\omega)-2\omega
\zeta(a)]$, and $\omega=K$, $\omega' = i K'$ are the real and imaginary
half-periods of the Jacobi elliptic functions, $\sigma$ and $\zeta$ are
the non-elliptic Weierstrass functions, and the factorization energy
$\alpha$ and $a$ are related by
\begin{equation}
\alpha = \frac23 (m+1) -\wp (a), \label{fe-d}
\end{equation} 
where $\wp$ is the well known Weierstrass function \cite{WV}. Thus:
\begin{eqnarray}
& u^\beta(x)u^{1/\beta}(x+\delta) =  
\frac{\sigma^2(x_0+\omega')}{\sigma(x_0+a+\omega')\sigma(x_0-a+\omega')}
\frac{\sigma(x+a+\omega')\sigma(x-a+\delta +\omega')}{
\sigma(x+\omega')\sigma(x+\delta+\omega')} e^{\delta\zeta(a)}
\end{eqnarray}
By taking $a=\delta$ we arrive at:
\begin{equation}
\hskip-1cm u^\beta(x)u^{1/\beta}(x+\delta) =
\frac{\sigma^2(x_0+\omega')e^{\delta\zeta(\delta)}}{\sigma(x_0+\delta 
+\omega')\sigma(x_0-\delta +\omega')} 
\end{equation}
where the right hand side does not depend on $x$, as was to be proved. It
is not difficult to check that our criterion is not satisfied for the
Lam\'e functions with $n>1$ (\cite{WV}, Section 23.7). 

Let us remind that the former results permitted to induce $\delta = T/2$
\cite{DF,KS,FNN}. This is now recovered for $\alpha = E_0$ and for the
unique Bloch eigenfunction $u(x) = \psi_0(x)$ satisfying $\psi_0(x)
\psi_0(x \pm T/2) = c$.

\medskip
\begin{figure}[htbp]
\centering \epsfig{file=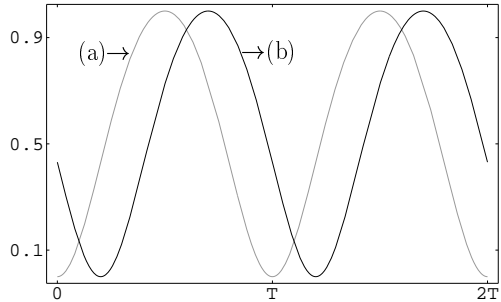, width=6.5cm}
\caption{\footnotesize The translational effect of the 2-nd order Darboux
transformation. (a) The initial Lam\'e potential with $n=1$ and $m=0.5$; 
(b) The 2-susy equivalent. The factorization energies $\alpha_1=1.1,$ 
$\alpha_2=1.4$ belong to the first energy gap $(E_1,E_{1'})$, the
displacement $\delta''=0.747\neq T/2$. The final effect is very simple but
it is not reducible to the nonsingular 1-st order steps.}
\end{figure}

\medskip

The existence of the corresponding second order displacements in the upper
spectral gap follows from the inverse spectral theorems. Indeed, if the
transformation (\ref{Wr}) involves the Bloch functions
(\ref{bloch},\ref{blochit}), it does not change the spectrum of $V_0$. As
the shape of the Lam\'e potential with $n=1$ is uniquely defined by its
band structure (cf. the inverse scattering theorems \cite{HOC,novi,BBEI},
see also \cite{RS} p. 299, Theorem XIII.91.b), we infer that the
non-singular Darboux transformation of Proposition~1 can cause no more but
a coordinate displacement. Once again, the abstract argument can be
supported by the explicit formulae. In fact, by superposing a pair of
Darboux operations (\ref{DV}) with the complex transformation functions
$u_i(x)$ given by (\ref{blochlame}), where $a=\delta_i$ are the complex
displacements $\delta_1 = \delta + \omega'$, $\delta_2 = \delta'
+\omega'$, one returns to the original Lam\'e potential displaced by
$\delta'' = \delta + \delta'$ (Fig.~6). The corresponding second order
transformation (\ref{Wr}) has the real Wronskian: 
\begin{equation}
W(u_1,u_2) = W_0 \frac{\sigma(x + \delta_1 + \delta_2 +
\omega')}{\sigma(x + \omega')} e^{-[\zeta(\delta_1)+ \zeta(\delta_2)](x -
x_0)}, \label{wlame}
\end{equation}
where 
\begin{equation}
W_0 = \frac{\sigma(\delta_2 - \delta_1) \sigma^2(x_0 +
\omega')}{\sigma(\delta_1)\sigma(\delta_2)\sigma(x_0 + \delta_1 + \omega') 
\sigma(x_0 + \delta_2 + \omega')}. \label{wini}
\end{equation}

\medskip 
\begin{figure}[htbp] 
\centering
\epsfig{file=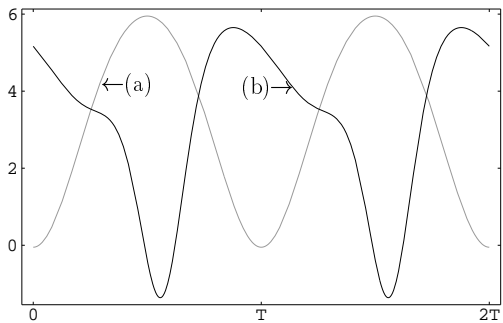, width=6.5cm} 
\caption{\footnotesize A non-trivial result of the periodicity preserving
second order Darboux transformation.  (a) The original Lam\'e potential
with $n=3$ and $m=0.5$ (b) The Darboux deformed version.  The
factorization energies $\alpha_1=2.15$, $\alpha_2=4.05$ belong to the
energy gap $(E_1,E_{1'})$. The global displacement affects the minima and
maxima but simultaneously the potential is deformed.} 
\end{figure}

\medskip

To examine the limitations of the method we have applied the second order
Darboux transformations \cite{SMPLA,S,sta,F,RO,bagchi} to the case $n=3$
(see Fig.~7).  A simple comparison with previous results \cite{FNN} shows
that the effect is global but essentially new (one sees the displaced
minima and maxima but also a non-trivial deformation). As it is already
known, the first order Darboux transformations cannot displace the Lam\'e
potentials with $n>1$ (see \cite{FMRS}). The problem whether the same
effect can be produced by higher order Darboux operations is still open
(compare the discussion of Khare and Sukhatme \cite{KS} with Dunne and
Feinberg \cite{DF}).

Once clarified the mechanism of the Darboux invariance, we have used the
transformation functions in (\ref{DV}-{\ref{Wr}) defined as nontrivial
linear combinations of the Bloch basis. Notice that the corresponding
Darboux operations must affect the periodic structure, since on both
extremes $x\rightarrow \pm \infty$ the transformation function reduces to
two different Bloch functions, causing two opposite asymptotic
displacements. On Fig.8 we show the periodicity defect and the injected
localized state of the Lam\'e potential (\ref{lame}) with $n=1$, due to
the 1-st order transformation (\ref{DV}) with $\alpha <E_0$. The Fig.9, in
turn, shows a defect of the same potential caused by the 2-nd order
(irreducible) transformation (\ref{Wr}), which has injected a pair of
localized states at the factorization constants $\alpha_1=1.2$, $\alpha_2
= 1.3$. In both cases, the effect {\it looks local} but it is not (in
fact, it is enough to compare minima and maxima of the initial and
transformed potentials). The resulting periodicity defects arise as if a
detail of the lattice was crushed by two opposite supersymmetric
displacements, creating a ``Darboux model'' for the contact effects. 

\medskip
\begin{figure}[htbp]
\centering
\hskip-0.5cm\epsfig{file=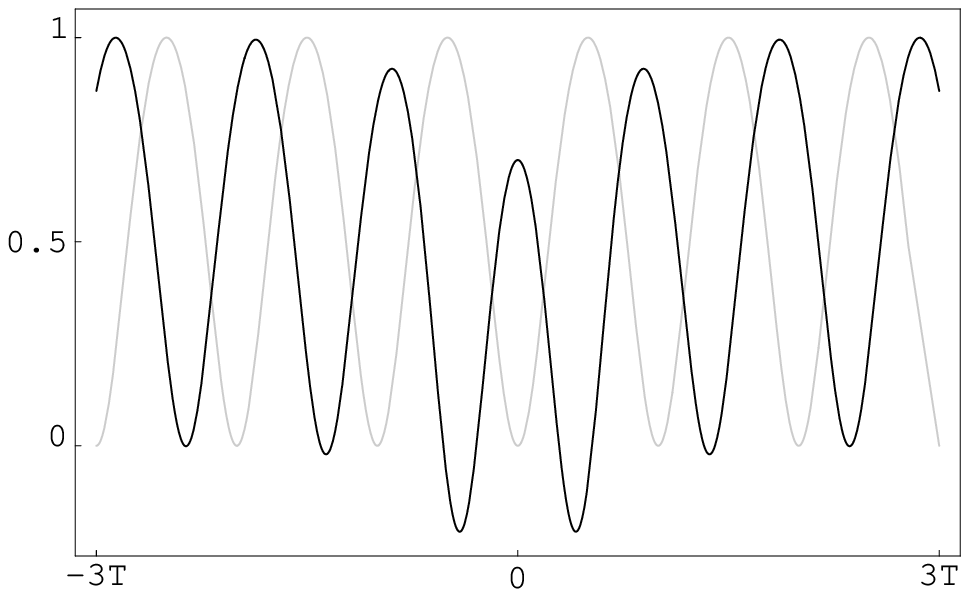, width=6.5cm} \hskip1cm
\epsfig{file=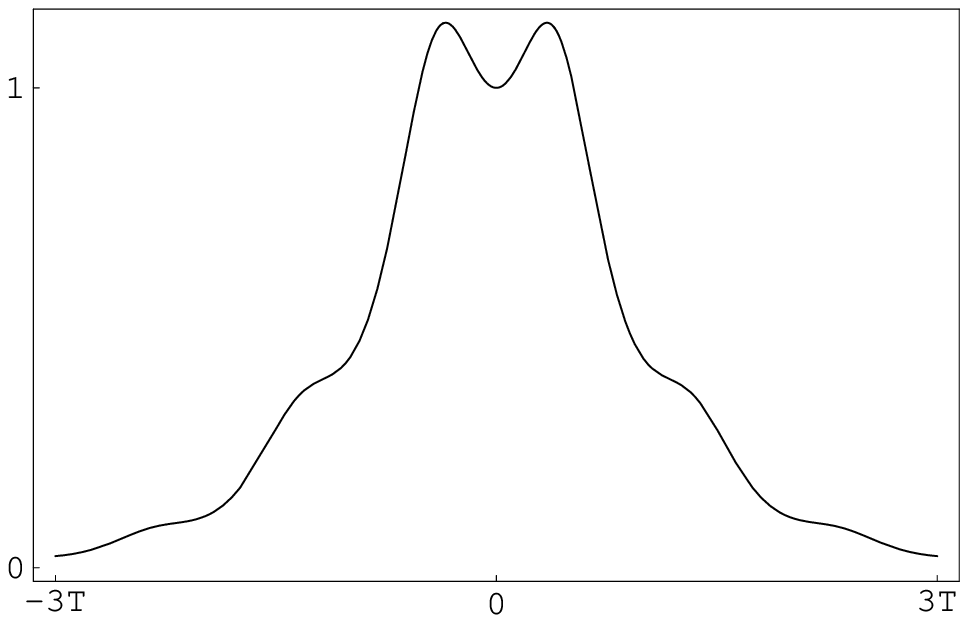, width=6.5cm}
\centerline{(a) \hskip6.7cm (b)}
\caption{\footnotesize The result of the first order Darboux
transformation applied to the Lam\'e potential with $n=1$ and $m=0.5$: (a) 
the supersymmetrically generated periodicity defect. Notice the asymptotic
translational invariance; (b) the energy bound state for $\alpha=0.35$
injected into the infinite forbidden band $(-\infty,E_{0})$.}
\end{figure}
\medskip

\medskip
\begin{figure}
\centering
\hskip-0.5cm\epsfig{file=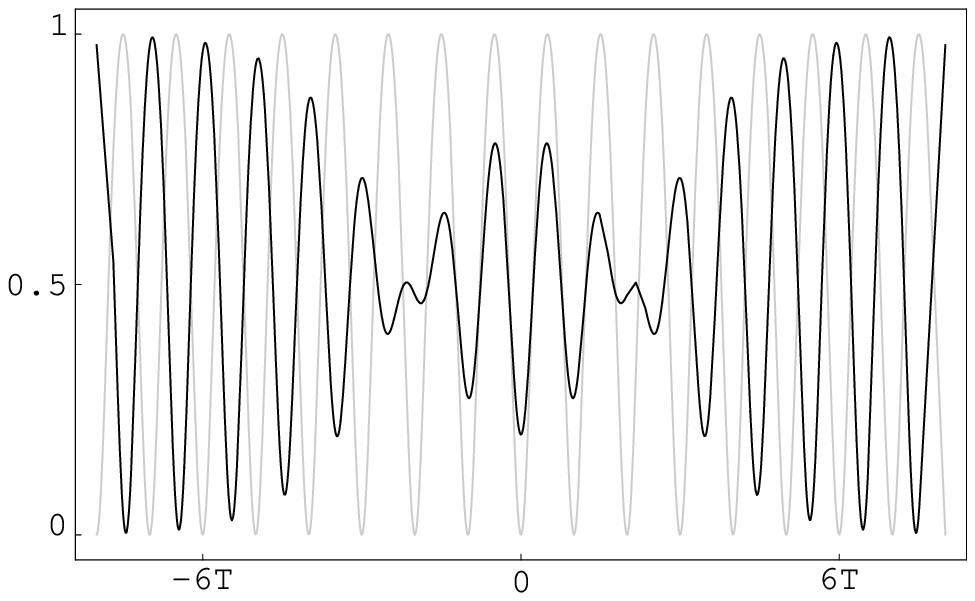, width=6.5cm} \hskip1cm
\epsfig{file=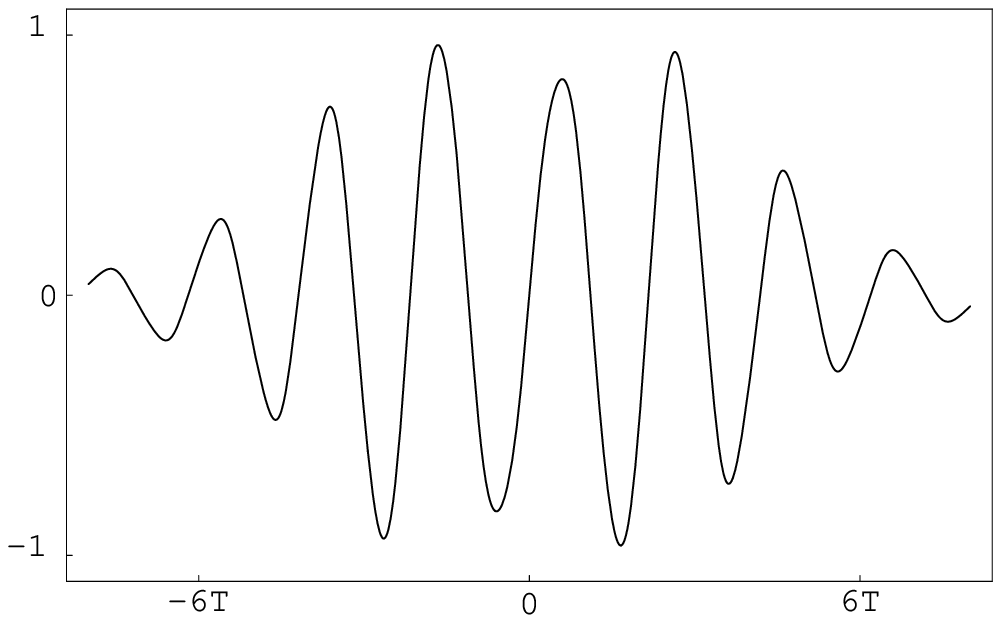, width=6.5cm}
\centerline{(a) \hskip6.7cm (b)}
\caption{\footnotesize The result of the second order Darboux
transformation (\ref{Wr}) with the factorization constants $\alpha_1 =
1.2, \ \alpha_2=1.3$ applied to the Lam\'e potential with $n=1$ and
$m=0.5$:  (a) the supersymmetrically generated periodicity defect.  Notice
the asymptotic translational invariance; (b) one of the energy bound
states for $\alpha_1=1.2$ injected into the forbidden band
$(E_1,E_{1'})$.}
\end{figure}
\medskip

To summarize, let us underline again an intriguing role of the
translational invariance, which can appear either as an exact or
approximate and/or asymptotic effect. This is an exceptional situation
when the most elementary symmetry transformation, the unitary but {\it
nonlocal} finite displacements, can be implemented by Darboux
transformations typically producing {\it local} but {\it non-unitary}
effects, a phenomenon specially convenient for the exact description of
the contact effects (tentatively, including tiny quantum wells
\cite{dhht}).

\newpage
\noindent
{\bf Acknowledgements.} The authors acknowledge the support of CONACYT
project 32086E (M\'exico).  BFS is grateful for the support of the Russian
Foundation for Basic Research and for the kind hospitality at the Physics
Department of CINVESTAV in the spring of 2001. ORO acknowledges the
support of CINVESTAV, project JIRA'2001/17.

\bigskip


\end{document}